\documentclass[aps,prl,reprint,superscriptaddress,nofootinbib]{revtex4-2}

\usepackage{amsmath,amssymb,bm}
\usepackage{amsthm}
\usepackage{mathtools}
\usepackage{combelow}
\usepackage{changes}
\usepackage{graphicx}
\usepackage[colorlinks=true,linkcolor=blue,citecolor=blue,urlcolor=blue]{hyperref}
\usepackage{microtype}
\usepackage{color}
\usepackage[nolist]{acronym}

\newcommand{\argmin}{\operatorname*{arg\,min}}
\newcommand{\mcA}{\mathcal A}

\newcommand{\mcZ}{\mathcal{Z}}
\newcommand{\dd}{\mathrm{d}}
\newcommand{\ket}[1]{\left|#1\right\rangle}
\newcommand{\bra}[1]{\left\langle #1\right|}
\newcommand{\braket}[2]{\left\langle #1|#2\right\rangle}

\newcommand{\old}{\color{black}}

\theoremstyle{definition}

\begin{document}

\title{Thermalization packets and optimal ice cubes}

\author{Israel Klich}
\email{ik3j@virginia.edu}
\affiliation{Department of Physics, University of Virginia, Charlottesville, VA 22904, USA}
\affiliation{Max Planck Institute for the Physics of Complex Systems, Dresden, Germany}

\author{Marija Vucelja}
\email{mv8h@virginia.edu}
\affiliation{Department of Physics, University of Virginia, Charlottesville, VA 22904, USA}
\affiliation{Max Planck Institute for the Physics of Complex Systems, Dresden, Germany}
\affiliation{Department of Mathematics, University of Virginia, Charlottesville, VA 22904, USA}

\date{\today}

\begin{abstract} 
Relaxation toward equilibrium is usually accelerated by modifying the environment or by cooling a system further from equilibrium. Here we introduce a distinct strategy: thermalization packets, auxiliary systems prepared in advance and later coupled to a target system to accelerate its relaxation toward a prescribed thermal state. Thus, thermalization packets trade preparation effort for reduced waiting time. When the objective is cooling, we colloquially refer to such packets as ice cubes. Unlike ordinary coolants, thermalization packets are characterized not only by their temperature or heat capacity, but also by their microscopic preparation. We define perfect and optimal packets by their ability to suppress the slowest relaxation mode of the coupled dynamics: perfect packets eliminate it entirely, while optimal packets minimize its amplitude. We show that, under generic conditions, perfect packets exist among thermal preparations near equilibrium. Surprisingly, for asymptotic relaxation, the optimum among thermal preparations is generally not the coldest packet: cooling the packet beyond the slow-mode-cancelling optimum restores a nonzero slow mode and can therefore slow relaxation, yielding a packet analog of the Mpemba effect. We demonstrate the concept in exactly solvable Metropolis dynamics, a minimal two-qubit model, and a boundary-coupled interacting Ising-spin system. We also extend the framework to packets optimized for finite readout times. Finally, we show that the perfect-packet contour can connect the trivial bath-equilibrium point to a nontrivial strong Mpemba or strong inverse-Mpemba point.
\end{abstract}

\maketitle

\begin{acronym}
    \acro{ME}{Mpemba effect}
    \acro{DoF}{degree of freedom}
    \acro{SM}{Supplemental Material}
    \acro{PDF}{probability density function}
    \acro{1D}{one-dimensional}
    \acro{2D}{two-dimensional}
    \acro{KMS}{Kubo–Martin–Schwinger}
    \acro{RP}{relaxation packet}
    \acro{TP}{thermalization packet}
\end{acronym}

\begin{figure}
    \centering
    \includegraphics[width=\columnwidth]{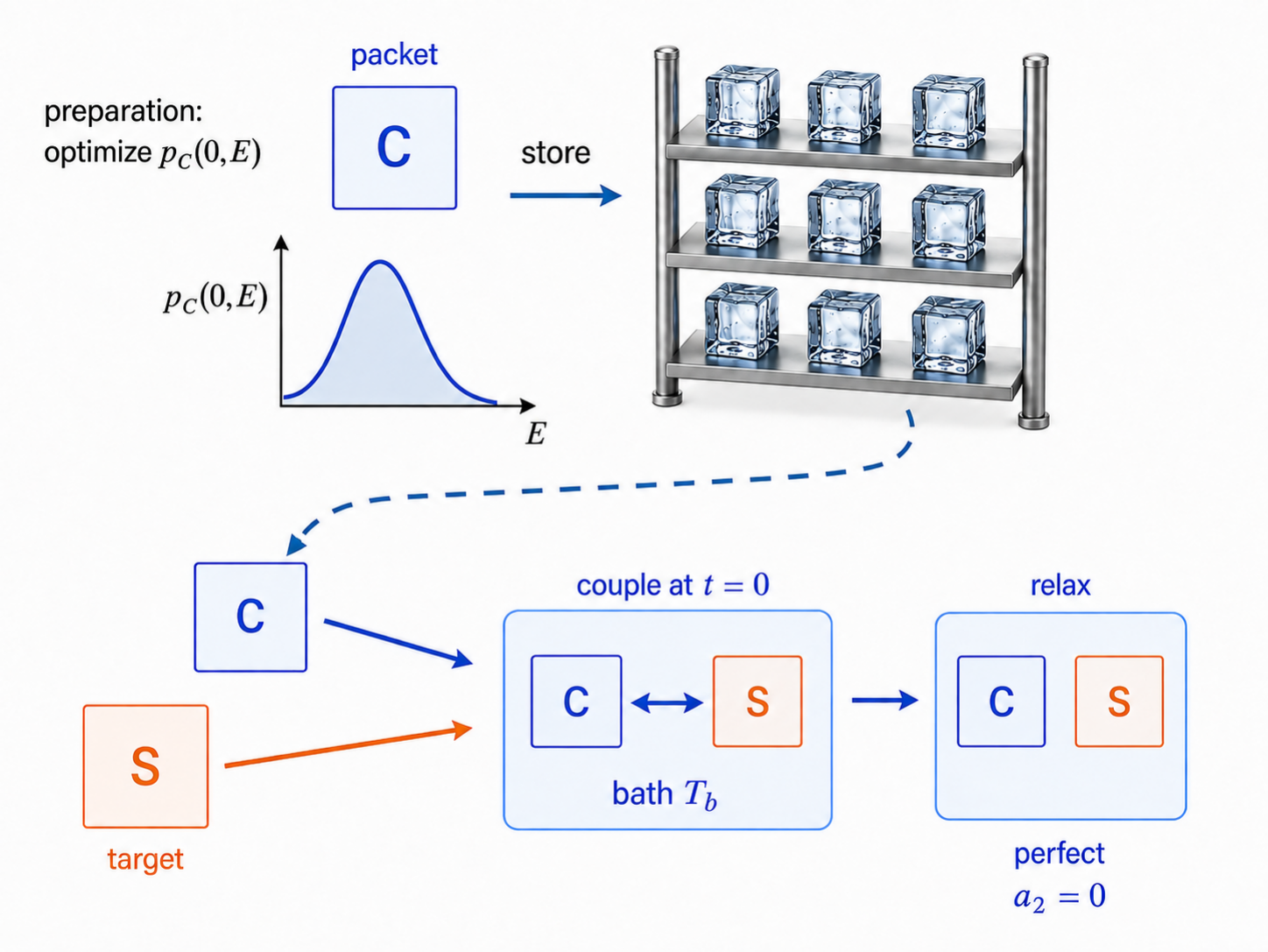}
    \caption{\textbf{Relaxation packets.} A finite auxiliary system $C$ is prepared independently from a target system $S$. At time $t=0$ the systems are coupled to each other and to a bath at $T_b$. The preparation $\ket{p_C(0)}$ affects the slow-mode amplitude $a_2$ of the post-coupling dynamics. A \emph{perfect packet} has $a_2=0$. In cooling applications, we refer to the packet as an \emph{ice cube}. 
    }
    \label{fig:fig-scheme}
\end{figure}

Cooling is usually described as the removal of heat by contact with a colder body, a flowing coolant, or a refrigerant in a thermodynamic cycle. This language emphasizes temperature, heat capacity, latent heat, thermal conductivity, and energy flow. Here, we ask a more microscopic question: how should an auxiliary system be prepared in advance if the goal is not only to absorb energy, but to make a target system relax rapidly? We call such a system a {\it relaxation packet}: an auxiliary system prepared to accelerate relaxation toward an arbitrary prescribed stationary state in the presence of a joint bath. Throughout the present work, we restrict attention to thermal stationary states, and accordingly use the term {\it thermalization packet}; in cooling applications, we colloquially call it an \emph{ice cube}\footnote{The analogy with an ordinary ice cube should not be taken too literally. A conventional ice cube is typically used to cool a system below the ambient temperature for a finite time rather than to reach a bath temperature. In this sense, it is more naturally viewed as a special case of the broader notion of a relaxation packet.}. The situation is schematically described in Fig.~\ref{fig:fig-scheme}. The central idea is that an auxiliary coolant should be designed to minimize --- or ideally cancel --- slow dynamical modes, rather than simply to maximize heat extraction or minimize its temperature.

Throughout this work, we regard time as the primary resource to be optimized. Specifically, we seek to minimize the time required for a target system to reach thermal equilibrium once the protocol begins. We assume that thermalization packets are prepared in advance and are available on demand. Their preparation time and energetic cost are therefore treated as sunk costs, analogous to preparing ice cubes or batteries before they are used. Under this assumption, the optimization problem concerns only the subsequent relaxation dynamics. Thus, thermalization packets trade preparation effort for reduced waiting time. Many auxiliary cooling protocols use the auxiliary as an entropy sink or an engineered reservoir; here, we instead optimize its preparation spectrally to eliminate slow modes while leaving the final equilibrium state unchanged.

We are particularly interested in perfect thermalization packets, whose preparation eliminates the slowest relaxation mode of the combined dynamics. More generally, an optimal packet minimizes the amplitude of this mode within the allowed class of preparations. We show that, under generic conditions, when the slowest relaxation mode is nondegenerate, such perfect packets can be chosen as Boltzmann states in the vicinity of thermal equilibrium. Surprisingly, making the packet as cold as possible generally destroys this perfect cancellation and results in slower relaxation of the combined system. Thus, unlike an ordinary ice cube, the optimal thermalization packet is not necessarily the coldest one, but rather the one whose preparation matches the target system. This counterintuitive behavior constitutes a ``packet'' analog of the Mpemba effect. 

In spectral terms, a Mpemba effect occurs when a state farther from equilibrium has a smaller overlap with the slowest relaxation mode than a state initially closer to equilibrium. Historical accounts note anomalous cooling behavior of this kind in water~\cite{aristotle1923meterologica}, as do more recent observations by Mpemba and Osborne~\cite{mpemba1969cool}, which sparked modern scientific interest. Mpemba and related phenomena occur in a variety of physical systems~\cite{lu2016nonequilibrium,Klich2019,Lasanta2017,KlichVucelja2026,kumar2020exponentially,KumarChetriteBechhoefer2022,Carollo2021,nava2019lindbladb,murciano2024entanglement,ares2025quantum,walker2021anomalous,walker2022mpemba,holtzman2022landaua,degunther2022anomalous,busiello2021inducing,schwarzendahl2022anomalous,zhang2022theoretical,teza2023relaxation,biswas2020mpembaa,santos2020mpemba,chetrite2021metastablea,shapira2024inverse,biswas2023mpembaa,summer2025resource,chittari2023geometric, shapira2026extendingmpembaeffectunderdamped,gonzales2021slow,Chatterjee2024universal,Carollo2021,baity-jesi2023memorya,nava2019lindbladb,yang2020nonmarkovian} and have recently attracted substantial scientific interest, both for their fundamental significance and practical implications.   

\emph{Spectral formulation and definitions ---} Let the target system and packet, after coupling, evolve on a state space with a continuous-time generator $W$, obeying
\begin{align}
\label{eq:probevolution}
    \partial _t \ket{p(t)} = W \ket{p(t)}\,.
\end{align}
We consider cases where the system relaxes to a stationary distribution and the generator is diagonalizable. The spectral solution of Eq.~\eqref{eq:probevolution} is  
\begin{align}
    \label{eq:spectral}
    \ket{p(t)}=\ket{v_1}+\sum_{k\ge2}a_k e^{\lambda_k t}\ket{v_k},
\end{align}
where $\lambda_k$ are the ordered eigenvalues 
$0=\lambda_1>\mathrm{Re}(\lambda_2)\ge\mathrm{Re}(\lambda_3)\ge\cdots$, $a_k$ are the projections of the initial condition $\ket{p(0)}$ on the left eigenvectors $\bra{u_k}$, 
\begin{align}
    a_k = \frac{\braket{u_k}{p(0)}}{\braket{u_k}{v_k}}\,, \label{eq:ak}
\end{align} 
while $\ket{v_k}$ are the corresponding right eigenvectors. We fix the normalization of the right eigenvectors throughout. The stationary distribution is the right eigenvector corresponding to the $\lambda _1 = 0$ eigenvalue, $\ket{v_1}$. In the presence of a spectral gap, the asymptotic relaxation time is controlled by the amplitude of the slowest mode. For example, in a case with $\mathrm{Re}(\lambda_2) > \mathrm{Re}(\lambda_3)$, a system starting with an initial condition with no projection onto $\bra{u_2}$, i.e. $a_2=0$, relaxes asymptotically at a rate of $-\mathrm{Re}(\lambda_3)$ or faster.

{\bf Definitions:} \emph{Thermalization packet, Perfect packet, Boltzmann packet} --- \emph{A thermalization packet} is an auxiliary system $C$, prepared independently of the target system $S$, and then coupled to $S$ and a joint bath with the objective of accelerating relaxation toward a prescribed thermal state. In the Markovian context, we say that a packet is \emph{perfect} if the initial state after coupling is such that the projection onto the eigenspace of the slowest modes is zero analogous to the condition for a strong Mpemba effect introduced in~\cite{Klich2019}. 
For a single slowest mode, the condition is simply 
\begin{align}
    \label{eq:perfect}
    a_2=0\,.
\end{align}
A \emph{Boltzmann packet} has as an initial condition the Boltzmann distribution at a temperature $T_p$,
\begin{align}
\pi_C(y;T_p)\equiv \frac{1}{\mcZ_C(T_p)}e^{-\frac{E_C(y)}{k_BT_p}}\,,
\end{align}
where $y\in \Omega_C$ is a microstate of the packet with corresponding energy $E_C(y)$, $\mcZ_C(T_p)$ is the partition function, 
\begin{align}
   \mcZ_C(T_p) = \sum _{y \in \Omega _C}e^{-\frac{E_C(y)}{k_BT_p}}\,, 
\end{align}
and $k_B$ is the Boltzmann constant, which we set to unity below. A packet may be engineered with any admissible probability distribution or, in quantum mechanics, with any admissible density matrix. Given an allowed set of preparations  $\mcA$, an optimal packet satisfies
\begin{align}
    \ket{p_C^\star(0)}&\in\argmin_{p_C(0)\in\mcA}|a_2(\ket{p_C(0)})|\,. \label{eq:optimal}
\end{align}
Perfect packets are the special case in which the minimum is zero. 

In the rest of the paper, we restrict attention to optimal packets for \emph{bath-mediated dynamics}. We assume no direct interaction between the packet and the target system, but rather interaction mediated by the bath. In this case, one can write 
\begin{align}
    \label{eq:additiveenergies}
    E(x,y)= E_S (x) + E_C(y)\,,
\end{align}
where $E(x,y)$ is the energy of the combined system-plus-packet in state $(x,y) \in \Omega _S \times \Omega _C$, $E_S(x)$ is the energy of the system in system state $x\in\Omega_S$, and $E_C(y)$ is the energy of the packet in packet state $y\in \Omega_C$. In this case, the system-plus-packet Boltzmann distribution factorizes,
\begin{align}
    \label{eq:jointBoltzmann}
    \ket{\pi_S(T_b)}\otimes\ket{\pi_C(T_b)}\,,
    %\pi(x,y;T_b)=\pi_S(x;T_b)\pi_C(y;T_b)\,,
\end{align}
where $T_b$ is the bath temperature. 

\begin{figure}
    \centering
\includegraphics[width=0.95\columnwidth]{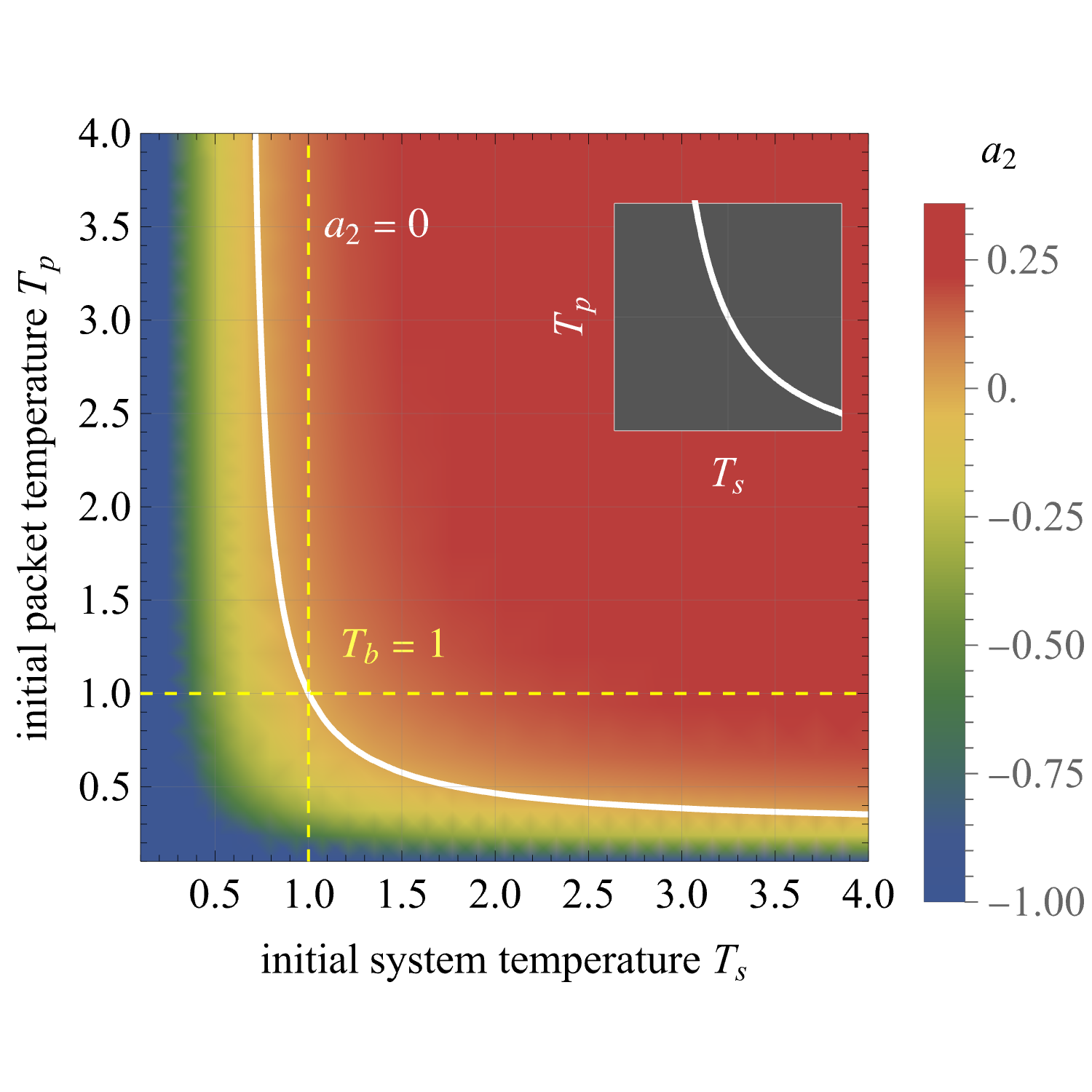}
    \caption{\textbf{Complete-graph Metropolis packet.} The color scale shows the projection $a_2$, defined in Eq.~\eqref{eq:MHslow}, for two independently Boltzmann-prepared subsystems: target system at temperature $T_s$ and a packet at temperature $T_p$. The target system and packet have $10$ and $5$  levels respectively, defined as follows: $E_S(i) = \epsilon i$ with $i= 1,\ldots, 10$, and $E_C(j) = \epsilon j$ with $j = 1,\ldots,5$. The bath temperature is fixed at $T_b=1$ and $\epsilon = 0.1 T_b$. The white curve denotes the perfect-packet condition $a_2=0$ [Eq.~\eqref{eq:MHperfect}] and passes through the equilibrium point $(T_s,T_p)=(T_b,T_b)$. The inset shows a magnified view of the $a_2=0$ contour near equilibrium, illustrating the generic local structure described by Eq.~\eqref{eq:localcurve}. Because $a_2$ is monotonic in each temperature separately for this model, the resulting curve represents a tunable compensation between the target and packet preparations, rather than a conventional strong Mpemba effect.
  \old}
    \label{fig:MH}
\end{figure}

\emph{Existence of perfect Boltzmann packets in bath-mediated thermalization} --- Boltzmann packets are more restrictive. They trace only the curve $\ket{\pi_C(T_p)}$ parametrized by $T_p$ in the space of probability distributions of the packet. A \emph{perfect Boltzmann packet} exists if $a_2(T_p)=0$ for an allowed $T_p$. The above expression implies a local existence result. Consider 
\begin{align}
    f(T_s,T_p)=a_2\left[\ket{\pi_S(T_s)}\otimes\ket{\pi_C(T_p)}\right]\,.
\end{align}
Note that for bath-mediated dynamics $f(T_b,T_b)=0$, as the combined system is already thermalized in this case. Thus, the implicit function theorem gives a local perfect-Boltzmann-packet curve as
\begin{align}
    f(T_s,T_p^*(T_s))=0\,,
\end{align}
with
\begin{align}
    \left.\frac{\dd T_p^*}{\dd T_s}\right|_{T_b}
    =
    -\frac{\partial_{T_s}f}{\partial_{T_p}f}
    \bigg|_{(T_b,T_b)}\,,
    \label{eq:localcurve}
\end{align}
provided $\left.\partial_{T_p}f\right|_{(T_b,T_b)}\ne 0$. 
Therefore, perfect Boltzmann packets are generically present near equilibrium. In temperature variables, the curve passes through $(T_b, T_b)$; for cooling, it pairs a nearby hot target with a colder packet when the local slopes have the appropriate signs. An example of the $a_2=0$ curve in a Metropolis dynamics model is presented below and illustrated in Fig.~\ref{fig:MH}.

It is worth noting that the statement can fail when the slow mode is insensitive to the packet coordinate, or when the slow eigenspace is multidimensional, in which case the scalar condition $a_2=0$ is replaced by the simultaneous cancellation of several slow-mode amplitudes. In that case, the perfect-packet condition has higher codimension, and a single tunable packet temperature will generically be insufficient to cancel all slow-mode amplitudes; isolated solutions may nevertheless occur.

\emph{Thermal bath coupling} --- Let $\alpha = (x,y)$ and $\alpha'=(x',y')$ be system-plus-packet states, with energies specified in Eq.~\eqref{eq:additiveenergies}. We choose a nonnegative symmetric connectivity $\Gamma_{\alpha',\alpha}=\Gamma_{\alpha,\alpha'}$ and transition rates 
\begin{align}
    W_{\alpha\to \alpha'}&\equiv W_{\alpha',\alpha}=\Gamma_{\alpha',\alpha}\,g\left(\Delta E_{\alpha',\alpha}\right)\,,
\end{align}
where $\Delta E_{\alpha',\alpha} \equiv E(\alpha') - E(\alpha)$ is the energy difference, and the diagonal is chosen so that $\sum_{\alpha'}W_{\alpha',\alpha}=0$, ensuring probability conservation. We assume that the ratio of the rates obeys \emph{detailed balance}, 
\begin{align}
    \label{eq:genericDB}
    \frac{W_{\alpha',\alpha}}{W_{\alpha,\alpha'}}&= \frac{g(\Delta E_{\alpha',\alpha})}{g(\Delta E_{\alpha,\alpha'})} = e^{-\frac{E(\alpha') - E(\alpha)}{T_b}}\,.
\end{align}
Equation ~\eqref{eq:genericDB} ensures detailed balance with respect to the factorized thermal state in Eq.~\eqref{eq:jointBoltzmann}. Despite the factorized thermal state, Eq.~\eqref{eq:jointBoltzmann}, the connectivity may nevertheless be nonseparable, as 
\begin{align}
\Gamma_{xy,x'y'}=\Gamma^S_{xx'}\delta_{yy'}+\Gamma^C_{yy'}\delta_{xx'}+\Gamma^{SC}_{xy,x'y'}\,,
\end{align}
with connectivity $\Gamma^{SC}$ allowing joint moves, where both the state of the system and the packet change simultaneously, i.e., $x\ne x'$ and $y\ne y'$. Note that $\Gamma^{S}$, $\Gamma ^C$ and $\Gamma^{SC}$ are symmetric. When the condition \eqref{eq:additiveenergies} holds, we call such a coupling a \emph{bath-mediated kinetic coupling}: equilibrium thermodynamics remains noninteracting, while the relaxation eigenmodes become joint. The examples below instantiate such couplings through exchange moves, boundary bath-mediated spin updates, and complete-graph Metropolis moves.

We now turn to examples to illustrate thermalization packets and their optimization. 

\emph{Complete-graph Metropolis packet} --- In Fig.~\ref{fig:MH} we give a first illustration of a perfect Boltzmann packet in the context of Metropolis dynamics. Consider the complete-graph Metropolis generator with stationary distribution $\ket{\pi(T_b)}$ \cite{Metropolis1953},
\begin{align}
\label{eq:Mrate}
W_{\alpha,\alpha'}=\begin{cases}
\dfrac{1}{n}\min\!\left(1,\dfrac{\pi(\alpha;T_b)}{\pi(\alpha';T_b)}\right),&\alpha\ne \alpha',\\
-\sum_{\delta\ne \alpha'}W_{\delta,\alpha'}, & \alpha = \alpha',
\end{cases}
\end{align}
where $n$ is the number of states of the joint system. We take $\pi(\alpha;T_b)$ to be the Boltzmann distribution of the joint system, Eq.~\eqref{eq:jointBoltzmann}. Remarkably, the eigensystem of this dynamics on the complete graph is known exactly \cite{Liu1996,KlichVucelja2026}. We partition the state space into a target system $S$ and a packet $C$ with additive energies, according to Eq.~\eqref{eq:additiveenergies}.

We prepare the joint system in the product state
\begin{align}
\label{eq:initMetropolis}
p(\alpha,0)=\pi_S(x;T_s)\pi_C(y;T_p)\,.
\end{align}

We denote with $x_1$ and $y_1$ the states minimizing the system and packet energies, respectively, so that $\alpha_1 = (x_1,y_1)$ is the joint ground state. The exact slow-mode coefficient is
\begin{align}
a_2(T_s,T_p;T_b)=\frac{\pi(\alpha_1;T_b)}
{1-\pi(\alpha_1;T_b)}
\left[
1-
\frac{p(\alpha_1,0)}
{\pi(\alpha_1;T_b)}
\right]\,.
\label{eq:MHslow}
\end{align}
A perfect Boltzmann packet is, therefore, obtained whenever
\begin{align}
p(\alpha_1,0)=\pi(\alpha_1;T_b)\,,
\end{align}
which, upon substituting Eqs.~\eqref{eq:initMetropolis} and~\eqref{eq:jointBoltzmann}, becomes
\begin{align}
\pi_S(x_1;T_s)\pi_C(y_1;T_p)
=
\pi_S(x_1;T_b)\pi_C(y_1;T_b)\,.
\label{eq:MHperfect}
\end{align}
For fixed bath temperature, this equation defines a contour in the $(T_s,T_p)$ plane, shown in Fig.~\ref{fig:MH}. From the cooling perspective, the packet temperature can be tuned so that the bath temperature lies between the initial temperatures of the target and packet subsystems, allowing the packet to compensate for the target system's excess slow-mode amplitude.

The exact solution also shows why this contour should not be identified with a conventional one-parameter strong Mpemba effect. Differentiating Eq.~\eqref{eq:MHslow} yields
\begin{align}
\operatorname{sgn}(\partial_{T_s}a_2)
=
\operatorname{sgn}\!\left(
\langle E_S\rangle_{\ket{\pi_S(T_s)}}-E_S(x_1)
\right),
\label{eq:MHmono}
\end{align}
which is positive unless the target spectrum is completely degenerate; the analogous result holds for $T_p$. Consequently, the condition in Eq.~\eqref{eq:MHperfect} expresses a compensation between two independently prepared subsystems rather than a sign change induced by varying a single control parameter. In the packet framework, however, this is precisely the desired mechanism: partitioning alone need not produce a conventional Mpemba effect, yet it can eliminate the slowest relaxation mode of the combined system.

\begin{figure*}
    \centering
    \includegraphics[width=0.98\textwidth]{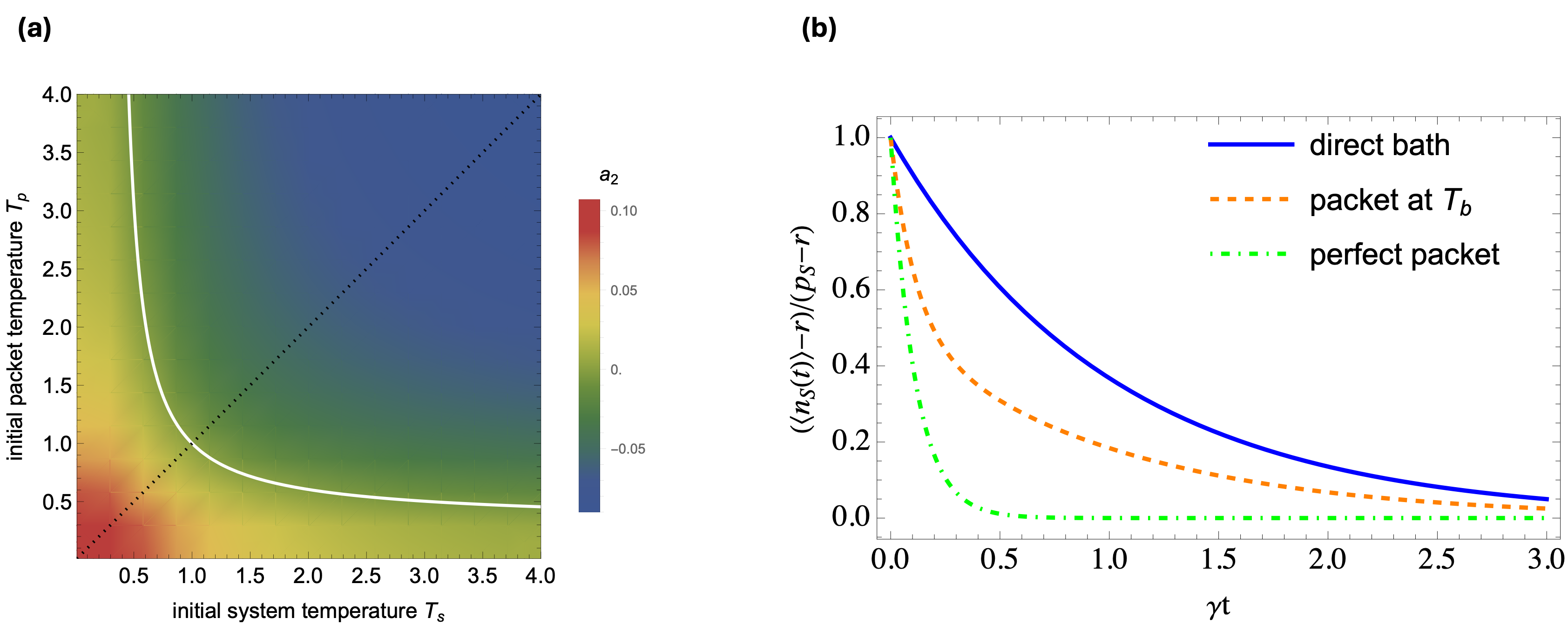}
    \caption{
    \textbf{Two-qubit example.}  (a) {Slowest mode amplitude.} The color scale shows the projection $a_2$, see Eq.~\eqref{eq:a2-qubit-perfect}, for two independently Boltzmann-prepared subsystems at temperatures $T_s$ and $T_p$. The bath temperature is fixed at $T_b/\epsilon=1$. The white curve denotes the perfect-packet condition $a_2=0$ [Eq.~\eqref{eq:qubit-perfect}] and passes through the equilibrium point $(T_s,T_p)=(T_b,T_b)$. (b)  The plot shows the relaxation of the normalized excited fraction excess of the target system, $(\langle n_S(t)\rangle-r)/(p_S-r)$, for $\kappa = 0$, i.e., ``direct'' cooling (blue solid), and two curves for $\kappa/\gamma=4$ --- one with a packet initially at the bath temperature (orange dashed), and the other with a \emph{perfect packet} (green dot-dashed). The target system is initially at infinite temperature with $p_S=1/2$. A perfect Boltzmann packet has $p_C^*=2r-p_S=0.0379$, or $T_p^*= 0.309\,\epsilon$, and removes the slow $e^{-\gamma t}$ contribution to the target excitation. }
    \label{fig:2qubitNoStrong}
\end{figure*}

\emph{Minimal two-qubit realization ---} As the simplest example, let both the target system and the packet be qubits with Hamiltonian
\begin{align}
\mathcal{H}&=\epsilon(n_S+n_C)\,,
\end{align}
and bath excited-state population 
\begin{align}
r&=\frac{1}{1+e^{\beta _b\epsilon}}\,,
\end{align}
where $\beta_b = 1/T_b$ is the inverse bath temperature. Each qubit undergoes thermal excitation $0\to1$ at rate $\gamma r$ and relaxation $1\to0$ at rate $\gamma(1-r)$, corresponding to a standard Davies thermal generator~\cite{Davies1974,Davies1976,BreuerPetruccione}. Coupling the qubits additionally allows the energy-conserving exchange process $10\rightleftarrows01$ at rate $\kappa$.
The stationary state is a product of the Boltzmann states \eqref{eq:jointBoltzmann}. 
Besides the stationary state, the generator has a slow mode with eigenvalue $\lambda_2=-\gamma$, together with modes at eigenvalues $-2\gamma$ and $-\gamma-2\kappa$. The right eigenvector associated with the slow mode for this dynamics is given by:
\begin{align}
    -e^{\beta_b\epsilon}|00\rangle
   +\frac{1}{2} \left(e^{\beta_b\epsilon}-1\right)\left(|10\rangle
   +|01\rangle\right) +|11\rangle\,.
\end{align}
We prepare the target and packet with excited-state populations $p_S$ and $p_C$, i.e., with a probability vector 
\begin{align}
\left(\left(1-p_S\right)|0\rangle
   +p_S|1\rangle \right)\otimes
   \left(\left(1-p_C\right)|0\rangle
   +p_C|1\rangle \right)\,,
   \end{align} 
where 
\begin{align}
\label{eq:SCpopulations}
p_S=\frac{1}{1+e^{\beta_s\epsilon}}\quad\text{and}\quad p_C = \frac{1}{1 + e^{\beta_p\epsilon}}\,. 
\end{align}
Explicitly, via \eqref{eq:ak} we find the projection to the slow mode
\begin{align}
\label{eq:a2-qubit-perfect}
a_2=r\left(p_S+p_C-2 r\right)\,.
\end{align}
The perfect packet condition is therefore
\begin{align}
    \label{eq:qubit-perfect}
    p_C^*&=2r-p_S.
\end{align}
The results are exhibited in Fig. \ref{fig:2qubitNoStrong}(a). 
For cooling, $p_S>r$, so the perfect packet is colder than the bath, $p_C^*<r$. Whenever $0<p_C^*<1$, it is itself a Boltzmann state, with inverse temperature, 
\begin{align}
\beta_p^*\epsilon
&
=\ln\left(\frac{2 e^{  \beta
   _b\epsilon}-e^{\beta _s\epsilon}+e^{\left(\beta _b+\beta
   _s\right)\epsilon}}{1-e^{\beta _b\epsilon}+2
   e^{\beta _s\epsilon}}\right)\,,
\end{align}
where we used Eqs.~\eqref{eq:SCpopulations} and~\eqref{eq:qubit-perfect}. 

It is instructive to consider the expectation value of the system excitation. We find that it evolves as
\begin{align}
&\left\langle n_S(t)\right\rangle
   =p_{10}(t)+p_{11}(t),
\end{align}
where the explicit expression for the excess system excitation is 
\begin{align}
\nonumber
\left\langle n_S(t)\right\rangle -r 
   ={}&\frac{p_S+p_C-2
   r}{2} e^{-\gamma t }\text{ 
   } \\
   \label{eq:twotraj}
   &
   +\frac{1}{2} e^{-(\gamma + 2 \kappa
   )t}\left(p_S-p_C\right).
\end{align}

For perfect preparation, Eq.~\eqref{eq:twotraj} simplifies to
 \begin{align}
\langle n_S(t)\rangle-r
 ={}&
 (p_S-r) e^{-(\gamma+2\kappa)t},
 \end{align}
so the target relaxes at the exchange-enhanced rate $\gamma+2\kappa$ instead of the bath rate $\gamma$. Although the full joint distribution still contains faster two-body relaxation modes, the slow total-energy mode has been eliminated. This exponential acceleration is illustrated in Fig.~\ref{fig:2qubitNoStrong}(b).

Note that the same model admits a direct Lindblad realization. The diagonal sector of the density matrix reproduces the classical two-qubit dynamics, the exchange jumps conserve the total energy, and the stationary state is the product Gibbs state at temperature $T_b$. Consequently, the same preparation defines a perfect quantum packet. 

\begin{figure*}[t]
    \centering
    \includegraphics[width=0.98\textwidth]{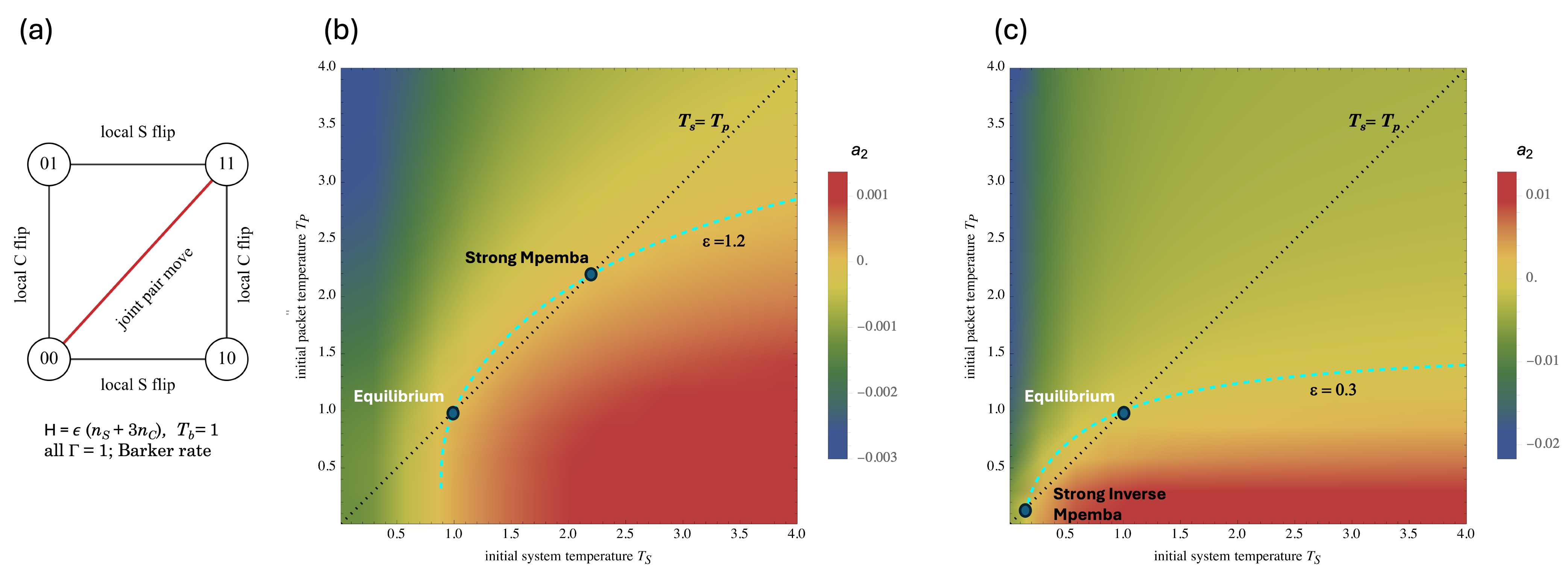}
    \caption{\textbf{Coexistence of a perfect-packet contour and a strong Mpemba point in two coupled qubits.}
     (a) The target qubit \(S\), with gap \(\epsilon\), and packet qubit \(C\), with gap \(3\epsilon\), are energetically noninteracting. The post-coupling dynamics contains local \(S\) and \(C\) flips and the bath-mediated joint pair move \(00\leftrightarrow11\); all attempt rates are \(\Gamma=1\), and Barker acceptance rates are evaluated at \(T_b=1\).
   (b,c) The slow-mode amplitude \(a_2\) for the factorized initial preparation \(\pi_S(T_s)\otimes\pi_C(T_p)\), for $\epsilon=1.2$ and $\epsilon=0.3$ . 
   The curves describe the perfect-packet condition \(a_2=0\), and the dotted line is \(T_s=T_p\). In each panel, one intersection is the equilibrium point (1,1); the second is a nontrivial strong Mpemba point in (b) and a strong inverse-Mpemba point in (c).}
    \label{fig:connected-strong-mpemba}
\end{figure*}

\emph{Finite-time optimization} --- In this section, we concentrate on the state of the system and average over the packet at a finite time. The condition in Eq.~\eqref{eq:qubit-perfect} eliminates the slow mode and is therefore optimal asymptotically. At a prescribed finite readout time \(t\), the packet and system are in general correlated, and a choice of colder packet may bring the target closer to equilibrium by allowing the slow and fast contributions in Eq. \eqref{eq:twotraj} to cancel at that time. Minimizing \(|\langle n_S(t_f)\rangle-r|\) over the physical packet population gives an optimal packet as a function of the finite readout time $t_f$
\begin{align}
p_{C,f}^*(t_f)
=
r-(p_S-r)\coth(\kappa t_f)\,,
\end{align}
provided the right-hand side is physically admissible. Otherwise, because the target deviation is linear in \(p_C\), the optimum lies at the nearest allowed boundary; for the cooling protocol considered here this is \(p_C=0\).

Since \(\coth(\kappa t_f)>1\), the finite-time optimal packet is colder than the asymptotically perfect packet. The latter is recovered continuously at long times,
\begin{align}
\lim_{t_f\rightarrow\infty}p_{C,f}^*(t_f)
=
2r-p_S
=
p_C^*\,.
\end{align}
For sufficiently early deadlines, the expression for \(p_{C,f}^*\) is negative, meaning that no physical packet can bring the target exactly to equilibrium by that time. The earliest exact equilibration is then obtained with the zero-temperature packet and occurs at
\begin{align}
t_{\min}
=
\frac{1}{\kappa}
\operatorname{arccoth}
\left(\frac{r}{p_S-r}\right)\,,
\end{align}
when \(r<p_S<2r\). For the parameters of Fig.~\ref{fig:2qubitNoStrong} (b), \(\gamma t_{\min}=0.3225\). Thus the slow-mode-canceling packet is the infinite-time limit of a family of deadline-matched packets, whereas finite-time optimization generally favors a colder preparation.

This fixed-time criterion allows transient arrival followed by overshoot. It should therefore be distinguished from the stricter problem of minimizing the time after which the target remains within a prescribed tolerance of equilibrium.

\emph{Perfect packet coexistence with the strong Mpemba effect} --- In the examples studied so far, the (strong) Mpemba effect is absent. In the strong Mpemba effect \cite{Klich2019}, the state of a system at thermal equilibrium at a temperature higher than the bath temperature has zero overlap with the slowest mode of the target bath. 

Here we consider a modification of the two-qubit model in which the $a_2 =0$ contour has an additional intersection with the equal-temperature line $T_s = T_p$. Depending on parameters, this intersection occurs above or below $T_b$, producing a strong Mpemba or strong inverse-Mpemba point. The results are illustrated in 
Fig.~\ref{fig:connected-strong-mpemba}. The target \(S\) and packet \(C\) are two initially decoupled qubits with additive Hamiltonian
\begin{align}
\mathcal{H}=\epsilon(n_S+3n_C)\,,
\end{align}
independently prepared in the product state \(\pi_S(T_s)\otimes\pi_C(T_p)\). At \(t=0\) they are kinetically coupled by allowing, in addition to the local flips of either qubit, the joint pair move \(00\leftrightarrow 11\). Every allowed transition \(\alpha\to\alpha'\) has attempt rate \(\Gamma_{\alpha,\alpha'}=1\) and Barker rate
\begin{align}
W_{\alpha'\leftarrow\alpha}
=\frac{\Gamma_{\alpha,\alpha'}}
{1+\exp\!\left[\beta_b\!\left(\Delta E_{\alpha',\alpha}\right)\right]}\,,
\end{align}
so the dynamics obeys detailed balance with respect to the equilibrium state at the bath temperature. Analyzing the system we find that the largest nonzero eigenvalue is 
\begin{align}
\lambda _2=-1\,.
\end{align}
The corresponding slow-mode amplitude is:
\begin{align}
    a_2=\frac{(-1+z) \left((-1+z) z+p_C
   \left(1+z^3\right)-p_S
   \left(1+z^3\right)\right)}{(1+z)^2 \left(1-3 z+7 z^2-8
   z^3+7 z^4-3 z^5+z^6\right)}\,,
\end{align}
where $z = e^{\beta_b \epsilon}$. The $a_2=0$ contour of perfect preparations is at \begin{align}
    p_C^*=\frac{p_S+z-z^2+p_S z^3}{1+z^3}.
\end{align}
At any fixed target temperature, its intersection with the corresponding vertical line selects a packet temperature that removes the slowest relaxation mode.

\begin{figure*}
    \centering
    \includegraphics[width=0.98\textwidth]{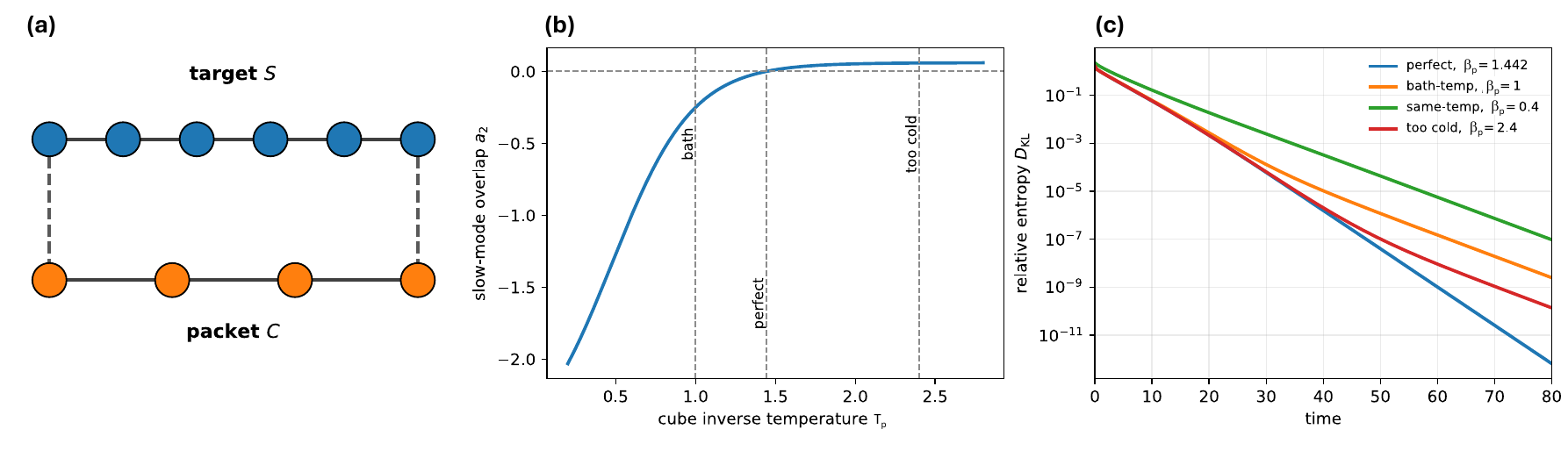}
    \caption{
    \textbf{Interacting $6+4$ spin ice cube.} \textbf{(a)} The target and packet have independent Ising Hamiltonians; they are dynamically coupled only by two bath-mediated boundary exchanges. \textbf{(b)} Exact diagonalization yields a zero of the slow-mode amplitude at $\beta_p^*=1.442\ldots$ for an initially hot target with $\beta_s=0.4$ and in a bath with $\beta_b=1$. \textbf{(c)} Exact relaxation of the relative entropy,  $D_\mathrm{KL}$. The perfect ice cube removes the tail associated with $\lambda_2$ and relaxes asymptotically with the tail associated with the $\lambda_3$ mode, while neutral or hot packets retain the slow mode.}
    \label{fig:spinpacket}
\end{figure*}

In contrast to Figs.~\ref{fig:MH}  and~\ref{fig:2qubitNoStrong}, where the $a_2=0$ contour intersects the equal-temperature line $T_s=T_p$ only at the bath-equilibrium point $T_s=T_p=T_b$, the contour in Fig.~\ref{fig:connected-strong-mpemba} has an additional intersection with this line. This additional intersection point is a genuine strong Mpemba point, where the system-plus-packet is initially prepared in Gibbs states at a common temperature, $T_s=T_p\neq T_b$, and the resulting joint state has zero overlap with the slow mode. Moreover, the curve switches behavior from strong Mpemba to strong inverse Mpemba as we change the energy scale $\epsilon$ (keeping $\beta_b=1$). Thus, the packet-optimization curve and the strong Mpemba effect are not competing mechanisms but two sections of the same zero-mode geometry; in particular, the optimal ``ice cube'' can itself be hotter than the bath.

\emph{Interacting spins} --- Here, going beyond the two-qubit example we demonstrate the acceleration in an interacting spin chain.  
Specifically, consider a target consisting of six Ising spins, $\{s_i\}$, and a packet of four Ising spins, $\{c_i\}$,
\begin{align}
\mathcal{H}_S&=-J_S\sum_{i=1}^{5}s_i s_{i+1}-h_S\sum_{i=1}^{6}s_i,\\
\mathcal{H}_C&=-J_C\sum_{j=1}^{3}c_j c_{j+1}-h_C\sum_{j=1}^{4}c_j,
\end{align}
with no energetic coupling between $S$ and $C$. Consequently, the equilibrium state is the product of the corresponding Boltzmann distributions. The dynamics consists of local Glauber spin flips within each chain together with bath-mediated boundary exchanges $10\rightleftarrows01$ connecting $(s_1,c_1)$ and $(s_6,c_4)$ at rate $\kappa$     (Fig.~\ref{fig:spinpacket}). Every transition is accepted with the Barker factor $(1+e^{\beta_b\Delta\mathcal{H}})^{-1}$, where $\Delta\mathcal{H}$ is the energy difference associated with the move. Detailed balance therefore holds with respect to the total Hamiltonian $\mathcal{H}=\mathcal{H}_S+\mathcal{H}_C$, even though the transition graph is not separable.

We use the parameters $J_S=0.75$, $h_S=0.70$, $J_C=0.90$, $h_C=0.50$, $T_b=1$, local flip rate $\gamma=0.36$, and boundary exchange rate $\kappa=6$ at each contact. The Markov chain contains $2^{10}=1024$ states and can be diagonalized exactly. Preparing the target hot, at $\beta_s=0.4$, and the packet in a Boltzmann state at inverse temperature $\beta_p$, the slow-mode amplitude $a_2(\beta_p)$ vanishes at
$\beta_p^*\simeq1.442\ldots$

Thus, a packet colder than the bath constitutes a perfect ice cube. The leading relaxation rates are $-\lambda_2=0.1021\ldots$ and $-\lambda_3=0.1837\ldots$, so eliminating the slowest mode accelerates the asymptotic relaxation by a factor $|\lambda_3|/|\lambda_2|\simeq1.80$. Figure~\ref{fig:spinpacket} shows both the zero of $a_2(\beta_p)$ and the relative entropy (Kullback--Leibler divergence),
\begin{align}
D_{\mathrm{KL}}\left(\ket{p(t)}\Vert \ket{\pi(T_b)}\right)
=
\sum_{\substack{s\in\Omega_S\\ c\in\Omega_C}}
p(s,c,t)
\ln\frac{p(s,c,t)}{\pi(s,c;T_b)}.
\end{align}
A packet prepared at the bath temperature, as well as a hot or overly cold packet, retains a nonzero $\lambda_2$ contribution to the state, whereas the perfect ice cube suppresses this contribution so that the leading deviation is governed by $\lambda_3$. Correspondingly, $D_\mathrm{KL}$ crosses over from an $e^{2\lambda_2t}$ to a $e^{2\lambda_3t}$ asymptotic decay. The curves are obtained by exact diagonalization of the finite-state Markov generator; the same dynamics can also be simulated using rejection-free kinetic Monte Carlo.

\emph{Relation to other auxiliary cooling protocols and outlook ---} 
Thermalization packets are related to, but distinct from, heat-bath algorithmic cooling, repeated-interaction models, reservoir engineering, and active-reset protocols, in which auxiliary systems often act as entropy sinks or engineered baths~\cite{Boykin2002,Baugh2005,RaeisiMosca2015,Alhambra2019,Ciccarello2022,Poyatos1996,Verstraete2009,Valenzuela2006}. A closely related recent qubit-reset protocol uses an incoherent ancilla and a single entangling gate to convert slowly decaying local coherences into faster-decaying global two-qubit coherences, thereby removing overlap with the slowest Liouvillian mode~\cite{Lejeune2026}. Our perspective is complementary: the auxiliary system itself is treated as a prepared thermodynamic resource, and the examples above suppress population-sector slow modes through packet preparation and bath-mediated kinetic coupling. The packet criterion, therefore, emphasizes the cancellation of slow modes rather than heat extraction alone. Optimal ice cubes generalize ordinary coolants into finite, prepared, spectral resources: the central question is no longer only how cold an auxiliary body is, but whether the allowed packet preparations can place the joint initial state on the fast manifold --- or, if not, how closely they can approach it under physical constraints.

In the present work, packet preparation is treated as a sunk cost. An important extension is to include preparation work and entropy production as explicit thermodynamic resources~\cite{Jarzynski2011}. More generally, one could optimize relaxation subject to constraints on preparation time, energy, or control complexity. When the target preparation is also controllable, one may similarly tune the target initial state to approach the fast manifold. At finite readout times, the optimum becomes horizon-dependent because faster modes remain relevant, suggesting a broader deadline-constrained control problem. Although we have emphasized cooling, the same construction applies to heating.

Experimental realizations would be particularly natural in platforms allowing independent auxiliary-state preparation and controlled joint dissipative dynamics, including qubit-based systems such as Refs.~\cite{shapira2024inverse, zhang2025observation} as well as classical stochastic systems. Analogous spectral-matching strategies in closed quantum systems would also be of interest~\cite{ares2025quantum,joshi2024observing} as well as connections to the control and calibration of quantum hardware~\cite{campaioli2026validation}. More broadly, thermalization packets provide a route to accelerating nonequilibrium relaxation by engineering finite auxiliary resources that are spectrally matched to the coupled dynamics, rather than by modifying the bath or the target system itself.

\begin{acknowledgments}
M.~V. and I.~K. acknowledge the stimulating environment and hospitality of MPI-PKS during the development of this work. The authors thank Beno\^it Dou\c{c}ot and  Gianluca Teza for engaging discussions. This material is based on work supported by the National Science Foundation under Grant
No.~DMR-1944539.
\end{acknowledgments} 

\bibliography{references}

\end{document}